\documentclass[prd,aps,eqsecnum,amsmath,floatfix,nofootinbib,preprint,tightenlines]{revtex4}

\usepackage{latexsym}
\usepackage{graphicx}
\usepackage{multirow}
\usepackage[dvipsnames]{xcolor}
\usepackage{hyperref}

\def\floatcaption#1#2{ \caption{#2 \label{#1}} }

\def\bibi{\bibitem}

\let\slo=\o                     

\def\c{\chi}

\def\e{\epsilon}                
\def\g{\gamma}

\def\j{\psi}

\def\l{\lambda}
\def\m{\mu}
\def\n{\nu}
\def\o{\omega}
\def\p{\pi}                     
\def\t{\tau}

\def\D{\Delta}

\def\G{\Gamma}

\def\S{\Sigma}

\def\cl{{\cal L}}
\def\cm{{\cal M}}

\def\cbo{{\,\raise-.15ex\Sc [\,}}                       
\def\ltap{\raisebox{-.4ex}{\rlap{$\sim$}} \raisebox{.4ex}{$<$}}   

\def\svev#1{\left\langle #1\right\rangle}       

\def\ddt#1{{\buildrel {\hbox{\LARGE .\kern-2pt.}} \over {#1}}}

\def\ie{\mbox{\it i.e.}}
\def\eg{\mbox{\it e.g.}}

\def\tr{{\rm tr}\,}

\def\half{{1\over 2}}

\def\ttl#1{{\it #1}}

\def\seef{{\it cf.\  }}

\def\hB{\hat{B}}

\def\hf{\hat{f}}

\def\bj{\overline{\j}}

\def\SU{{\rm SU}}

\begin{document}

\begin{boldmath}
\begin{center}
{\large{\bf
Application of dilaton chiral perturbation theory to $N_f=8$, $\SU(3)$ spectral data
}}\\[8mm]
Maarten Golterman,$^a$ Ethan T.~Neil$^b$ and Yigal Shamir$^c$\\[8 mm]
$^a$Department of Physics and Astronomy, San Francisco State University,\\
San Francisco, CA 94132, USA\\
$^b$Department of Physics, University of Colorado, Boulder, CO
80309, USA\\
$^c$Raymond and Beverly Sackler School of Physics and Astronomy,\\
Tel~Aviv University, 69978, Tel~Aviv, Israel\\[10mm]
\end{center}
\end{boldmath}

\begin{quotation}
We extend dilaton chiral perturbation theory (dChPT) to include the taste splittings in the Nambu--Goldstone sector observed in lattice simulations of near-conformal theories with staggered fermions.   We then apply
dChPT to a recent simulation by the LSD collaboration of the $\SU(3)$ gauge theory
with 8 fermions in the fundamental representation, which is
believed to exhibit near-conformal behavior in the infrared,
and in which a light singlet scalar state, nearly degenerate
with the pions, has been found.
We find that the mesonic sector of this theory can be
successfully described by dChPT, including, in particular,
the mesonic taste splittings found in the simulation.   We
confirm that current simulations of this theory are in the ``large-mass'' regime.
\end{quotation}

\newpage
\section{\label{intro} Introduction}
There has been a growing interest in the non-perturbative dynamics
of gauge theories with more light fermionic degrees of freedom than
QCD, obtained by increasing the number of fundamental fermions or by
taking fermions to be in larger representations, or both.   If all fermions
transform in a vector-like representation of the gauge group, such
theories can be studied on the lattice, and many groups have pursued
such simulations, both with an eye toward Beyond the Standard Model
(BSM) model building, and because the dynamics of such theories
may be qualitatively different from the dynamics of QCD.    For reviews of the lattice efforts,
we refer to Refs.~\cite{DeGrandreview,NP,Pica,Benlat17}.

An example of different dynamical behavior is the appearance in some of
these theories of a very light scalar with the same quantum numbers
as the very broad, and relatively heavy $f_0(500)$ resonance in QCD.
Specifically, in $\SU(3)$ gauge theory with either $N_f=8$ fundamental
Dirac fermions \cite{LSD,LSD2,LatKMI},\footnote{%
  For earlier work on the $N_f=8$ theory, see Ref.~\cite{LSD0}.
}
or two sextet Dirac fermions
\cite{sextetconn,sextet1,sextet2,Kutietal,Kutietal20},
a singlet $0^{++}$ scalar has been observed nearly degenerate in mass
with the ``pions,'' {\it i.e.}, the
pseudo-Nambu--Goldstone bosons (pNGB's) associated with chiral
symmetry breaking, at the fermion masses employed in these
simulations.   Similarly, a very light singlet $0^{++}$ scalar has
been observed in $\SU(3)$ gauge theory with four light and
eight heavier Dirac fermions in the fundamental representation
\cite{BHRWW}.
The appearance of the light singlet scalar in these simulations
is accompanied by the onset of approximate hyperscaling laws.
A similar behavior has also been reported recently in the $\SU(3)$ gauge theory
with four light and six heavier Dirac fermions in the
fundamental representation \cite{6plus4}.

Chiral perturbation theory (ChPT) has been a powerful tool for interpreting the results from simulations of lattice QCD.
In the case of theories with a light scalar, which in current
simulations is roughly degenerate with the pions, also the light scalar will have to be included in
an effective field theory (EFT) approach to interpreting the data.
Any such EFT should be constructed using the (approximate) symmetries of the underlying theory, and be based on a hypothesis for the parametrical smallness of the mass of the
light scalar, much like the usual assumption of chiral symmetry
breaking explains the smallness of the pion mass.
An EFT framework based on the
assumption that the light singlet scalar, which henceforth we will
refer to as the dilaton, can be viewed as a pNGB for approximate
scale invariance has been developed using a systematic spurion
analysis, and with a consistent power
counting, in Refs.~\cite{PP,latt16,gammay,largemass}.\footnote{Ref.~\cite{ApB}
already discussed some of the ideas underlying this construction.  For dChPT
with the pions in the $\e$-regime, see Ref.~\cite{BGKSS}.}
We will refer to this
framework as dilaton-ChPT, or dChPT for short.
dChPT is based on a systematic expansion in the fermion mass as well as
in the distance to the conformal window, as measured by the trace of
the energy-momentum tensor of the massless theory at the
chiral symmetry breaking scale.
For other approaches to include the light scalar in a low-energy description,
see Refs.~\cite{BLL,GGS,CM,MY,HLS,AIP1,AIP2,CT,LSDsigma,AIP3}.

In this paper, we fit lattice spectroscopy data from Ref.~\cite{LSD2}
for the $N_f=8$ $\SU(3)$ gauge theory to the predictions of tree-level
dChPT.   The simulations reported in Ref.~\cite{LSD2} were carried
out with $n$-HYP smeared staggered fermions, and exhibit taste splitting of the
pion multiplet (for reviews of taste breaking in QCD with
staggered fermions, see for instance Refs.~\cite{MILC,MGLH}).
The quantities we consider are the pion mass, the
dilaton mass, the pion decay constant, and the masses of two
non-singlet taste pions for which data are provided in Ref.~\cite{LSD2}.
We fit these quantities as a function of the bare fermion mass, taking correlations into account.
Fits of dChPT to the pion mass, the pion decay constant, and the dilaton mass have
been considered before in Ref.~\cite{Kutietal}, but dChPT fits to the taste-split pions and the inclusion of data correlations in the fits are new.
The behavior of the
taste-split pion spectrum as a function of the fermion mass is rather different from that in QCD, and
thus provides a particularly interesting way to test
dChPT, extended to include the effects of taste breaking.

In Sec.~\ref{dChPT} we briefly summarize dChPT at lowest order,
recasting predictions for masses and the pion decay constant in a
form that is useful for our fits.\footnote{Note that the dilaton decay
constant was not computed in Ref.~\cite{LSD2}.}   In Sec.~\ref{taste} we
analyze the effect of taste breaking associated with the use
of staggered fermions, and summarize expressions for the taste
breakings in the pion mutiplet, again in a form that is useful for our
fits.   Then, Sec.~\ref{fits} is concerned with the fits themselves,
after some preliminary remarks about the choice of units in which
to express the quantities to be fit.  Sec.~\ref{conclusion} contains
our conclusions, while an appendix discusses the use of the
gradient flow scale $t_0$.  Preliminary results have been presented in Ref.~\cite{MGlat19}.

\section{\label{dChPT} Tree-level dChPT}
The euclidean leading-order (LO) lagrangian for dChPT is given
by
\begin{eqnarray}
\label{LOlag}
\cl&=&\frac{1}{2}\hf_\t^2 e^{2\t}\partial_\m\t\partial_\m\t+
\frac{1}{4}\hf_\p^2 e^{2\t}\,\tr(\partial_\m\S^\dagger\partial_\m\S)\\
&&-\frac{1}{2}\hf_\p^2\hB_\p m e^{(3-\g_*)\t}\,\tr(\S+\S^\dagger)
+\hf_\t^2\hB_\t c_1 e^{4\t}\left(\t-\frac{1}{4}\right)\ .
\nonumber
\end{eqnarray}
Here $\hf_\t$, $\hf_\p$, $\hB_\p$ and $\hB_\t$ are low-energy
constants (LECs).   The dimensionless small parameter $c_1$ is proportional to
the small expansion parameter $n_f-n_f^*$, with $n_f$ defined as the
limiting value of $N_f/N_c$ in the Veneziano limit \cite{VZlimit},
where the number of colors $N_c$ and the number of fundamental flavors $N_f$
tend to infinity simultaneously.  $n_f^*$ is the value of $n_f$
for the theory at the conformal sill: the boundary between the regime
with chiral symmetry breaking and the regime in which the massless theory is
conformal in the infrared.     The effective field for the dilaton is
$\t$, and $\S=\mbox{exp}(2i\p/\hf_\p)$ is the usual non-linear field
describing the pion multiplet.   The $\t$ field has been shifted such that
$v(m) \equiv \svev{\t}=0$ for $m=0$.   Finally, because of the proximity of the sill of the conformal window $n_f=n_f^*$, where the gauge coupling $g$ runs into
the infrared fixed point $g_*$,
the value of the ``walking'' coupling is close to its value
at the infrared fixed point.  The same applies
to the mass anomalous dimension, $\g(g)$, which we can thus expand around $\g_*=\g(g_*)$, the mass anomalous dimension at the infrared fixed point at the conformal sill. For a detailed discussion of the
construction of this lagrangian, its relation to the underlying theory
with $N_f$ fundamental fermions and the power counting, we refer to
Refs.~\cite{PP,largemass}.

For $m>0$, the potential is minimized by $\S=1$, and  $v(m)$ then
solves the saddle-point equation
\begin{equation}
\label{saddle}
\frac{(3-\g_*)m}{4c_1\cm}=v(m)\,e^{(1+\g_*)v(m)}\ ,\qquad
\cm=\frac{\hf_\t^2\hB_\t}{\hf_\p^2\hB_\p N_f}\ .
\end{equation}
Furthermore, taking into account that the pion and dilaton fields need to be renormalized by a common factor $e^{v(m)}$, one obtains from Eq.~(\ref{LOlag})
\begin{subequations}
\label{masses}
\begin{eqnarray}
\label{massesa}
M_\p^2&=&2\hB_\p m e^{(1-\g_*)v(m)}\ ,\\
\label{massesb}
M_\t^2&=&4c_1\hB_\t e^{2v(m)}\left(1+(1+\g_*)v(m)\right)\ ,\\
\label{massesc}
F_\p&=&\hf_\p e^{v(m)}\ .
\end{eqnarray}
\end{subequations}

Next, we combine Eqs.~(\ref{massesa}) and~(\ref{massesc}),
using Eq.~(\ref{saddle}), to obtain
\begin{equation}
\label{comb}
\frac{M_\p^2}{F_\p^2}=\frac{8\hB_\p c_1\cm}{\hf_\p^2(3-\g_*)}\,v(m)\equiv \frac{1}{d_1}\,v(m)\ ,
\end{equation}
defining the constant $d_1$. dChPT is valid when
$M_\p^2/F_\p^2$ is parametrically small,
which is true as long as $c_1 v(m)$ is small enough.
First, when $\frac{m}{c_1\cm}\,\ltap\, 1$, also $v(m)\,\ltap\,1$, just leading
to the requirement that $c_1$ is small. Indeed, it is,
since $c_1\propto n_f-n_f^*$, which is small by assumption.   But, when
$\frac{m}{c_1\cm}\gg 1$, Eq.~(\ref{saddle}) implies that
\begin{equation}
\label{vapprox}
v(m)\sim \frac{1}{1+\g_*}\,\log\left(\frac{(3-\g_*)m}{4c_1\cm}\right)\ ,
\end{equation}
and the requirement that $M_\p^2/F_\p^2$ be parametrically small becomes
\begin{equation}
\label{lmregime}
c_1\log\left(\frac{m}{c_1\cm}\right)\ll 1\ .
\end{equation}
In the large-mass regime, \ie,
when $\frac{m}{c_1\cm}\gg 1$, using the approximate solution
Eq.~(\ref{vapprox}), we find that $M_\p$, $M_\t$ and $F_\p$ scale like
\begin{equation}
\label{hyperscaling}
M_\p\sim M_\t\sim F_\p\sim m^{\frac{1}{1+\g_*}}\ .
\end{equation}
This hypescaling behavior extends to other quantities as well
\cite{largemass}.  It can be understood by observing that
for $\frac{m}{c_1\cm}\gg 1$ the breaking of scale invariance
is dominated by the fermion mass $m$, instead of by the (slow) running
of the renormalized coupling.

We now return to Eq.~(\ref{saddle}), which we will solve exactly, \ie, we will not use the approximate solution~(\ref{vapprox})
in the rest of this paper.\footnote{
  It can be shown that Eq.~(\ref{vapprox}) is the leading term in an
  expansion of the exact classical solution in $\log m$
  and $\log\log m$.
  For details, see Ref.~\cite{largemass}.
}
We use Eq.~(\ref{comb}) to rewrite Eq.~(\ref{saddle}) as
\begin{equation}
\label{saddle2}
m=\frac{4c_1\cm}{3-\g_*}\,v(m)e^{(1+\g_*)v(m)}
=d_2\frac{M_\p^2}{F_\p^2}\,e^{(1+\g_*)d_1\frac{M_\p^2}{F_\p^2}}\ ,
\end{equation}
with
\begin{equation}
\label{D2}
d_2=\frac{4c_1\cm}{3-\g_*}d_1=\frac{\hf_\p^2}{2\hB_\p }\ .
\end{equation}
Eliminating $e^v$ from Eqs.~(\ref{massesa}) and~(\ref{massesc}), one finds
\begin{equation}
\label{mMpiFpi}
M_\p^2F_\p^{-1+\g_*}=2\hB_\p\hf_\p^{-1+\g_*} m
\equiv d_0 m\ ,
\end{equation}
so that
\begin{equation}
\label{Fpi}
F_\p=\left(\frac{d_0 m}{M_\p^2/F_\p^2}\right)^{\frac{1}{1+\g_*}}\ .
\end{equation}
The solution of Eq.~(\ref{saddle2}) for $M_\p^2/F_\p^2$ in terms of $m$
can be expressed using the Lambert $W$-function as\footnote{
  See, \eg, \tt{https://en.wikipedia.org/wiki/Lambert\_W\_function}.}
\begin{equation}
\label{solsaddle}
\frac{M_\p^2}{F_\p^2} = h(m) \equiv \frac{1}{(1+\g_*)d_1}\,W_0\left(\frac{(1+\g_*)d_1}{d_2}\,m\right)\ .
\end{equation}
This allows us to fit $M_\p^2/F_\p^2$ and $F_\p$ as well as
$M_\t^2/F_\p^2$ as functions of $m$:
\begin{subequations}
\label{Afit}
\begin{eqnarray}
\label{Afita}
\frac{M_\p^2}{F_\p^2}&=&h(m)\ ,\\
\label{Afitb}
F_\p&=&
\left(\frac{d_0 m}{h(m)}\right)^{\frac{1}{1+\g_*}}\ ,\\
\label{Afitc}
\frac{M_\t^2}{F_\p^2}&=&d_3\left(1+(1+\g_*)\,d_1 h(m)\right)\ ,\qquad d_3\equiv \frac{4 c_1 \hB_\t}{\hf_\p^2}
\ .
\end{eqnarray}
\end{subequations}
These are the equations we will fit in Sec.~\ref{fitAC}.
We note that both $1/d_1$ and $d_3$ are proportional
to $c_1$, which is parametrically small as a function of
the distance to the conformal window, $n_f-n_f^*$.  In this
paper, we will consider $n_f$ to be fixed, so that $c_1$ is
constant, and thus $d_1$ and $d_3$ are also constants.
In contrast, $d_0$ and $d_2$ are purely defined in terms
of LECs.

In the case of Eq.~(\ref{Afita}) it may be inconvenient to fit directly to $h(m)$
due to its dependence on the Lambert $W$-function.
Instead, one may then return to Eq.~(\ref{saddle2}) and carry out the fit
treating $m$ as a dependent variable.

\section{\label{taste} Extension to taste splittings}
Let us recall the scaling properties of the pion mass term
in dChPT.  One begins with the observation that,
under a scale transformation, $\bj\j \to \l^{3-\g_*}\bj\j$ at leading order in the dChPT expansion.
 This simple scaling relation holds
when $n_f$ is close to $n_f^*$, and we are at a scale which is close enough
to the chiral symmetry breaking scale.
  The scaling of $\bj\j$, in turn,
determines the scaling of the mass, $m\to \l^{1+\g_*}m$.  This leads to
the form of the pion mass term in the effective theory,
\begin{equation}
\label{pmass}
e^{(3-\g_*)\t} \hf_\p^2\hB_\p m\,\tr(\S+\S^\dagger) \ .
\end{equation}

A similar reasoning can be used to determine the structure of taste-breaking
operators in the leading-order effective lagrangian of a nearly conformal
theory.  We start from the Symanzik effective action, where the leading taste breaking
effects are encoded in four-fermion operators of the generic form
\cite{LS}
\begin{equation}
\label{Symstag}
a^2 (\bj \G \j) (\bj \G \j) \ ,
\end{equation}
where $a$ is the lattice spacing, and $\G$ stands for gamma matrices
that act on the taste index.  Under a scale transformation,
each of
these four-fermion operators will develop an anomalous dimension,\footnote{%
  Since the four-fermion operators transform in different representations of the lattice symmetry group, they do not mix under renormalization.
}
\begin{equation}
\label{scl4fermi}
(\bj \G \j) (\bj \G \j) \to \l^{6-\g_\G} (\bj \G \j) (\bj \G \j) \ ,
\end{equation}
where now $\g_\G$ is the value of the anomalous dimension at the
conformal sill.  Correspondingly,
we should treat $a^2$ as a spurion, transforming as
\begin{equation}
\label{scla}
a^2 \to \l^{-2+\g_\G} a^2 \ .
\end{equation}
Having fixed the transformation properties of $a^2$ (as a spurion
for this particular four-fermion operator), the corresponding expression
at the EFT level is
\begin{equation}
\label{chiralstag}
c_\G \hf_\p^6  a^2 e^{(6-\g_\G)\t} O_\G \ , \qquad
O_\G = \tr(\S \G \S^\dagger \G) \ ,
\end{equation}
where  $c_\G$ is a dimensionless LEC.
There are four different single-trace operators
which contribute to the tree-level mass splittings.\footnote{
  There are similar double-trace terms, which, however,
  contribute to the mass splittings only at the next order.
}
It follows that the mass squared of the pion with taste $\G'$ is
larger than the exact pNGB pion mass squared by an amount
\begin{equation}
\label{M2tst}
M^2_{\G'} - M_\p^2 =  \hf_\p^4 a^2 \sum_{\G}c'_{\G'\G} e^{(4-\g_{\G})v}\ .
\end{equation}
The ratios $c'_{\G'\G}/c_\G$ are pure numbers.
For the precise list of operators, and for the ratios $c'_{\G'\G}/c_\G$,
see Ref.~\cite{AB} (see also Ref.~\cite{MILC}).
For the pNGB pion of the exact chiral symmetry of the
massless staggered lattice action, the $c'$ coefficients all vanish.

Using Eq.~(19) of Ref.~\cite{AB}, and introducing
\begin{equation}
\label{EG}
  \D(\G_i) \equiv a^2(M_{\G_i}^2 - M_\p^2) \ , \qquad
  E(\g_i) = e^{(4-\g_i)v} \ ,
\end{equation}
one finds, for the tastes
\begin{equation}
\label{tastelist}
\G_i\in\{\G_5,\G_{\m 5},\G_{\m\n},\G_\m,\G_I\}\ ,
\end{equation}
the following tree-level mass splittings:\footnote{%
  The familiar tree-level QCD mass splittings would be recovered
  by setting $E(\g_i)=1$.
}
\begin{subequations}
\label{tsplit}
\begin{eqnarray}
\label{tsplit5}
  \D(\G_5) &\equiv& \D_P \ = \ 0 \ ,
\\
\label{tsplitA}
  \D(\G_{\m 5}) &\equiv& \D_A \ = \
  \phantom{1} C_1 E(\g_1) + 3 C_3 E(\g_3) + \phantom{1} C_4 E(\g_4) + 3 C_6 E(\g_6) \ ,
  \hspace{8ex}
\\
\label{tsplitT}
  \D(\G_{\m\n}) &\equiv& \D_T \ = \
  \hspace{12.5ex} 2 C_3 E(\g_3) + 2 C_4 E(\g_4) + 4 C_6 E(\g_6) \ ,
\\
\label{tsplitV}
  \D(\G_\m) &\equiv& \D_V \ = \
  \phantom{1} C_1 E(\g_1) + \phantom{1} C_3 E(\g_3) + 3 C_4 E(\g_4) + 3 C_6 E(\g_6) \ ,
\\
\label{tsplitS}
  \D(\G_I) &\equiv& \D_S \ = \
  \hspace{12.7ex} 4 C_3 E(\g_3) + 4 C_4 E(\g_4) \ .
\end{eqnarray}
\end{subequations}
We have absorbed $\hf_\p^4 a^2$ into the new constants $C_1$, $C_3$, $C_4$
and $C_6$.\footnote{%
  The coefficients $C_{1,3,4,6}$ are constant for the purpose of this paper,
  since the data of Ref.~\cite{LSD2} have been obtained at a common,
  fixed lattice spacing.
}
Values for the pion masses with tastes $\G_{\m 5}$ and $\G_{\m\n}$
have been reported in Ref.~\cite{LSD2}, and we will attempt to fit
$\D(\G_{\m 5})$ and $\D(\G_{\m\n})$ to these data.

\section{\label{fits} Fits to spectroscopic data}
The simulations of Ref.~\cite{LSD2} were done at 5 different fermion masses,
\begin{equation}
\label{fermionmasses}
am_i\in\{0.00125\,,\ 0.00222\,,\ 0.005\,,\ 0.0075\,,\ 0.00889\}\ ,
\end{equation}
all at the same bare coupling.  For the lattice spacing we adopt
a mass-independent prescription, where the lattice spacing
is taken to be a function of the bare coupling only.
Thus, we will assume that the lattice spacing is the same
for all 5 ensembles.
This assumption can be self-consistently tested, as will
do toward the end of Sec.~\ref{fitAC}.

The data for masses and decay constants in Ref.~\cite{LSD2} are all given
in units of $\sqrt{8t_0}$, with $t_0$ the flow parameter from the gradient
flow, which itself is computed in lattice units, as $\sqrt{8t_0}/a$.
We begin by converting the mean values of all dimensionful quantities
back to lattice units.
The covariance matrices of these data for each ensemble were provided to us
by the LSD collaboration, and all our fits are fully correlated.
Largely speaking, we find that correlations among these data are weak,
and have little effect on the results of the various fits presented below.
Furthermore, correlations between $t_0$ and all other quantities
were found to be so small that they can be neglected.

Our choice of units
begs the question as to why we do not express all dimensionful quantities
in units of $\sqrt{8t_0}$ before carrying out the fits.   In QCD, this
would be a natural approach, as $\sqrt{8t_0}$ is a quantity that can
be expressed in terms of the mass $m$ using ChPT \cite{BG}.
However, in the present case, because of the appearance of the
scaling factor $e^{v(m)}$ at tree level in all dimensionful quantities,
it turns out that no (useful)
chiral expansion for $\sqrt{8t_0}$ exists, as explained in App.~\ref{$t_0$}.

\begin{table}[t]
\begin{center}
\begin{tabular}{|c|c|c|}
\hline
 & 5 ens. & 4 ens.  \\
\hline
$\c^2$/dof & 11.9/10 & 2.9/7  \\
$p$-value & 0.29 & 0.89 \\
\hline
$\g_*$ & 0.933(19) & 0.936(19)\\
$\log{d_0}$ & 1.938(60) & 1.931(61)\\
$d_1$ & 0.250(26) & 0.232(23) \\
$\log{d_2}$ & -16.68(94) & -16.14(85) \\
$d_3$ & 2.83(31) & 3.03(32)\\
\hline
\end{tabular}
\end{center}
\floatcaption{tab:basic}{\it Results of fit to Eq.~(\ref{Afit}).
Middle column: Including all ensembles corresponding to the 5 mass values
in Eq.~(\ref{fermionmasses}).  Right column:
Omitting the highest-mass ensemble $(am=0.00889)$.
}
\end{table}

\subsection{\label{fitAC} Fit of the pion mass, the pion decay constant and the dilaton mass}
We begin with a $\c^2$ fit of the quantities in Eq.~(\ref{Afit}),
namely, $M_\p^2/F_\p^2$, $aF_\p$ and $M_\t^2/F_\p^2$, to data
for the 5 different fermion masses~(\ref{fermionmasses}).
We do not consider $F_\t$, as it was not measured in Ref.~\cite{LSD2}.
The fit to Eq.~(\ref{Afit}) contains 5 parameters for $3\times5=15$ data points,
and, therefore,
10 degrees of freedom.   We find $\c^2_{min}=11.9$, for a
$p$-value of 29\%.  We have determined the logarithms of the
parameters ${d_0}$ and $d_2$, instead of the parameters themselves,
as it turns out that this helps the stability of the fit.
The results of the fit
are shown in the middle column of Table~\ref{tab:basic}.
Errors are always computed by linear error propagation from the full data covariance matrix.
A fit not including the quantity $M_\t^2/F_\p^2$ yields virtually the same
parameter values and errors (except of course for $d_3$),
and a $p$-value of 11\%.

For reasons that will be discussed in Sec.~\ref{4ens} below,
we have also carried out a fit to data from 4 ensembles,
omitting the highest-mass $(am=0.00889)$ ensemble.
The $p$-value of this fit is 89\%, significantly larger
than that of the 5-ensemble fit.  The results of the 4-ensemble fit
are reported in the rightmost column of Table~\ref{tab:basic},
and plotted in Fig.~\ref{fitC}.  As can be seen in Table~\ref{tab:basic},
the results of the 5-ensemble and 4-ensemble fits are closely consistent with each other.

We take the result obtained in the 4-ensemble fit, $\g_*=0.936(19)$,
as our final result for the mass anomalous dimension.
All further 4-ensemble fits presented in the rest of this section
reproduce the same result for $\g_*$. The difference in central values
between the 4-ensemble and 5-ensemble fits could be taken as a measure
of the systematic uncertainty, but we note that this difference
is much smaller than the error obtained from the fit.

\begin{figure}[t!]
\vspace*{4ex}
\begin{center}
\includegraphics*[width=7cm]{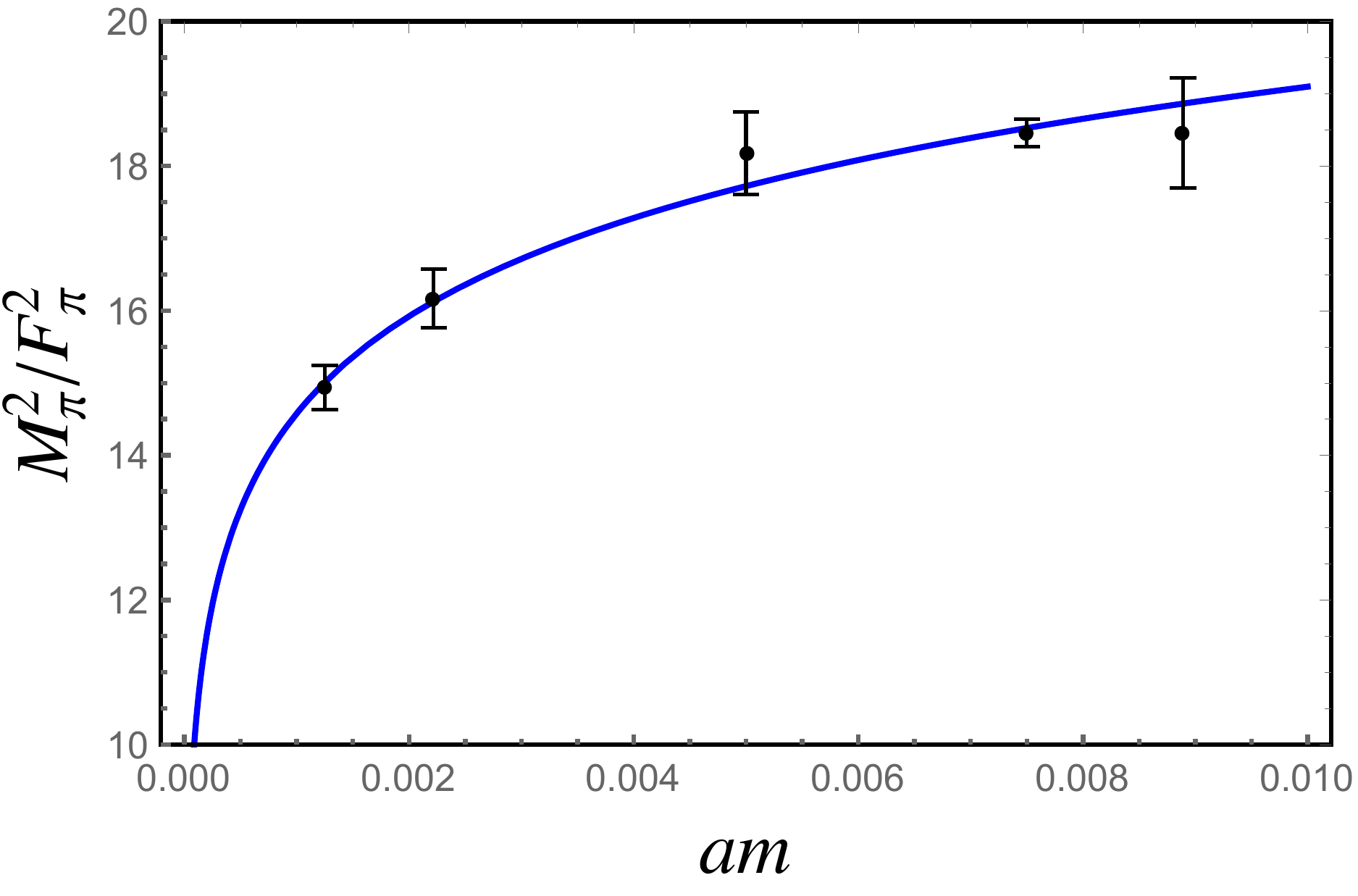}
\hspace{0.5cm}
\includegraphics*[width=7cm]{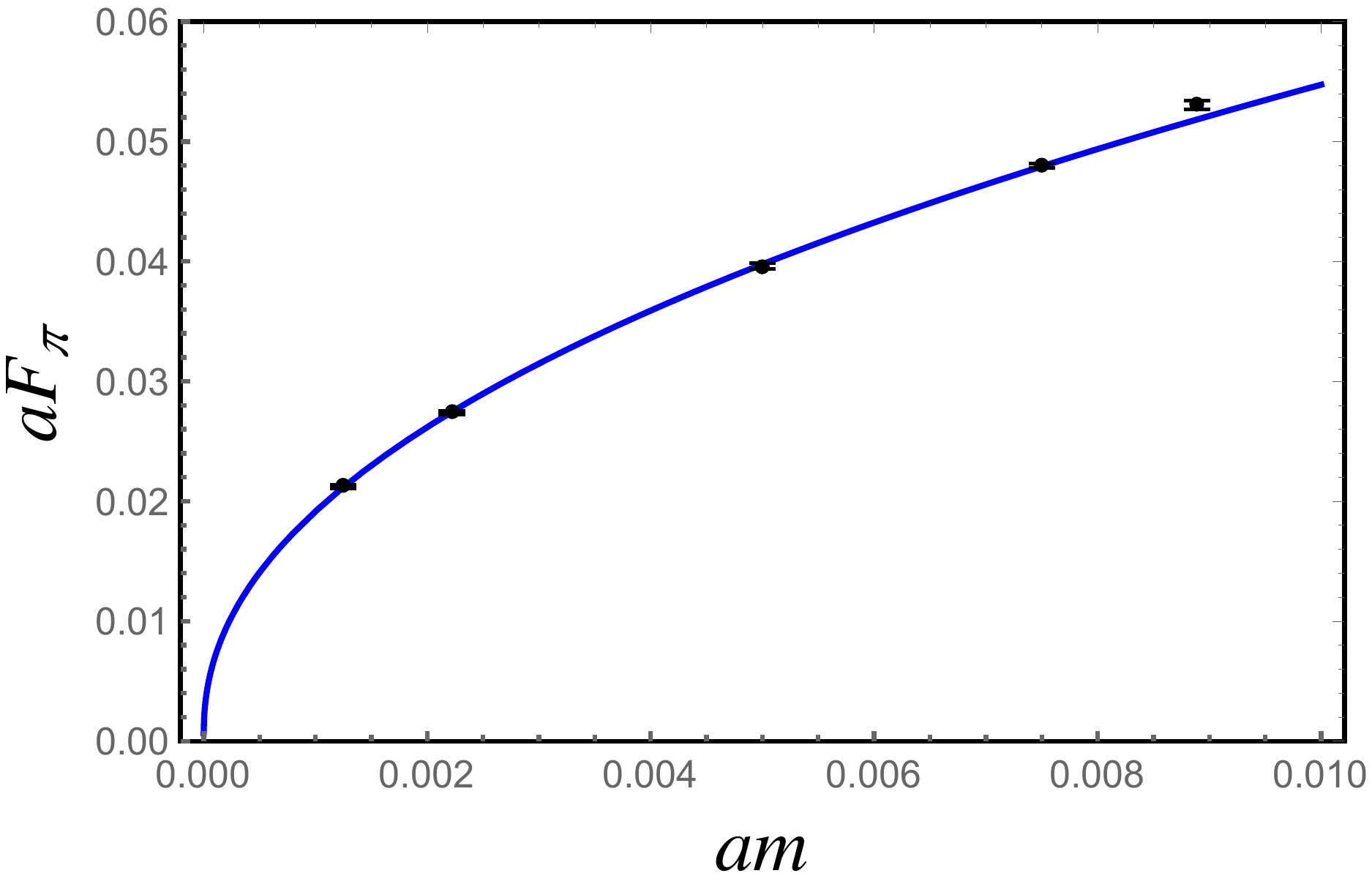}

\hspace{0.5cm}
\includegraphics*[width=7cm]{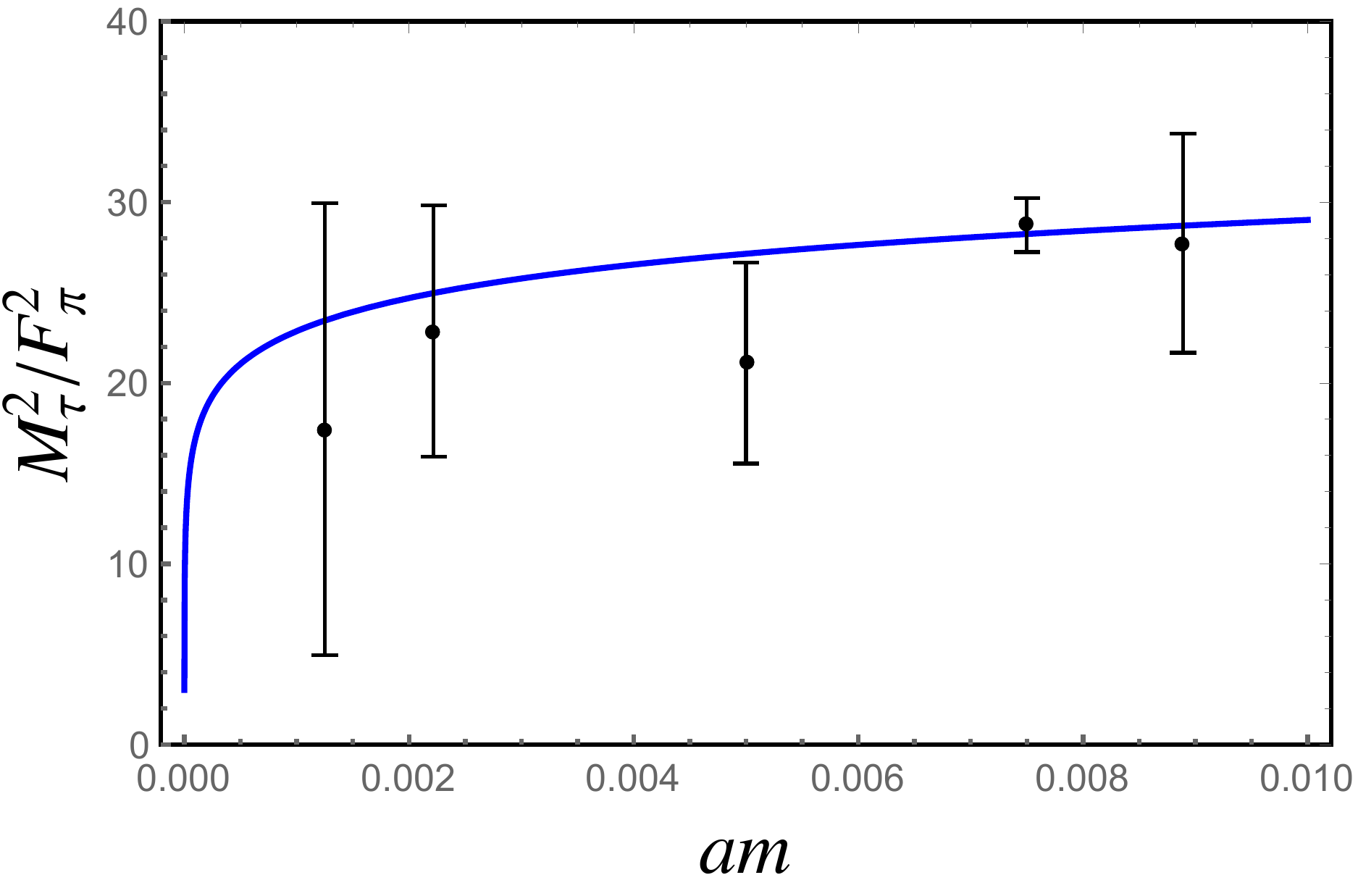}
\end{center}
\begin{quotation}
\floatcaption{fitC}%
{{\it Representation of the 4-ensemble fit reported in the
rightmost column of Table~\ref{tab:basic}.
The upper left-hand plot shows $M_\p^2/F_\p$ as a function of $am$,
the upper right-hand plot shows $aF_\p$,
and the lower plot shows $M_\t^2/F_\p^2$.}}
\end{quotation}
\vspace*{-4ex}
\end{figure}

Using the fit results from Table~\ref{tab:basic} we can infer
some additional tree-level parameters of dChPT, or combinations thereof.
The results for these derived quantities are collected
in Table~\ref{tab:derived}.  First,
\begin{eqnarray}
\label{fhatpi}
a\hf_\p&=&(d_0 d_2)^{\frac{1}{1+\g_*}}\ ,\\
a\hB_\p&=&\half\,d_0^{\frac{2}{1+\g_*}}d_2^{\frac{1-\g_*}{1+\g_*}} \ .
\nonumber
\end{eqnarray}
As can be seen from Tables~\ref{tab:basic} and~\ref{tab:derived},
the well-determined parameters are $\g_*$ and $\hB_\p$.
These are the parameters that control the mass dependence; whereas $\hf_\p$,
which has a much larger error, characterizes the massless theory.
The values we find for $\g_*$, $a\hf_\p$ and $a\hB_\p$ agree
well with the values found in Ref.~\cite{Kutietal}.\footnote{
  We note that Ref.~\cite{Kutietal} did not have access to the
  data correlation matrix.  The good agreement is in
  accordance with the fact that correlations are relatively weak.
}
The value of $\g_*$ is consistent with the earlier estimate of
Refs.~\cite{AIP1,AIP2}.
Using also our results for $d_3$ allows us to obtain the ratio of
the decay constants in the chiral limit, as well as the combination\footnote{%
  Because we have data at a single value of $n_f$,
  only the combination $c_1\hB_\t$ is accessible.
}
$c_1\hB_\t$ in units of $\hf_\p$,
\begin{eqnarray}
\label{fratio}
\frac{\hf_\t}{\hf_\p}&=&2\sqrt{\frac{3-\g_*}{d_1d_3}}\ ,\\
\frac{c_1\hB_\t}{\hf_\p^2}&=&\frac{1}{4}\,d_3\ .\nonumber
\end{eqnarray}
Combining these two expressions, we also have
\begin{equation}
\label{mdftaudivfpisq}
\frac{M_\t(m=0)\hf_\t}{\hf_\p^2}=2\,\frac{\sqrt{c_1\hB_\t}}{\hf_\p}\,\frac{\hf_\t}{\hf_\p}\ ,
\end{equation}
the value for which again is in good agreement with Ref.~\cite{Kutietal}.

\begin{table}[t]
\begin{center}
\begin{tabular}{|c|c|c|}
\hline
 & 5 ens. & 4 ens.  \\
\hline
$a\hf_\p$ & 0.00049(22) & 0.00065(27) \\
$a\hB_\p$ & 2.09(14) & 2.15(14) \\
$\hf_\t/\hf_\p$ & 3.415(86) & 3.427(88) \\
$c_1\hB_\t/\hf_\p^2$ & 0.708(77) & 0.757(81) \\
$M_\t(m=0)\hf_\t/\hf_\p^2$ & 5.75(32) & 5.96(32) \\
\hline
\end{tabular}
\end{center}
\floatcaption{tab:derived}{Derived quantities, see text.
}
\end{table}

To end this subsection, we return to the assumption that the
lattice spacing $a$ is independent of $am$.
Now that the fit parameters have been determined from a fit, we can test
this assumption self-consistently, with a precision set by the errors
of the fit.   In particular, the fit parameters allow us to
extract $a\hB_\p$ for each value of $am$ in the simulation separately,
using Eqs.~(\ref{massesa}) and~(\ref{comb}),
\begin{equation}
\label{Bhatpi}
a\hB_{\p,i}=\frac{(a M_{\p,i})^2}{2am_i}\,e^{(\g_*-1)d_1 M_{\p,i}^2/F_{\p,i}^2}\ .
\end{equation}
Since $\hB_\p$ is by construction independent
of $am$, this measures the dependence of $a$ on $am$.
Using Eq.~(\ref{Bhatpi}), and the results of the 4-ensemble fit,
we find the values
\begin{equation}
\label{Bhatpivalues}
a\hB_{\p,i}\ \in\
\{2.15(11)\,,\ 2.14(12)\,,\ 2.17(14)\,,\ 2.15(14)\,,\ 2.22(14)\}\ ,
\end{equation}
for each of the fermion masses~(\ref{fermionmasses}), respectively
(the values obtained from the 5-ensemble fit are very close).
In calculating the error in $\hB_{\p,i}$ we neglected the data errors for $M_\p$ and $F_\p$, and kept only the errors of
(and correlation between) $\g_*$ and $d_1$, since the latter are much larger.
We conclude that indeed $a\hB_\p$
is constant as a function of $am$, within errors.   The values
in Eq.~(\ref{Bhatpivalues}) are consistent with the extrapolated
value in Table~\ref{tab:derived}.

In principle, we could have used any of the (dimensionful) LECs
$a\hf_\p$, $a\hB_\p$, $a\hf_\t$ and $a^2 c_1 \hB_\t$ for this test.
We chose to use the one that is most precisely determined
from our fits, which is $a\hB_\p$.   The values of the three
other LECs are obtained by extrapolation to the chiral limit.
As can be seen in Table~\ref{tab:derived},
the precision of $\hf_\p$ is only 45\%, and this also sets
the precision with which we can determine $a\hf_\t$ and
$a^2c_1 \hB_\t$, \seef\ Eq.~(\ref{fratio}).

\begin{table}[t]
\begin{center}
\begin{tabular}{|c|c|c|c|}
\hline
$\c^2$/dof  & 16.0/12   & 19.8/14   & 19.8/14   \\
$p$-value   & 0.19       & 0.14       & 0.14       \\
\hline
$\g_*$      & 0.934(19)  & 0.932(19)  & 0.932(19)  \\
$\log{d_0}$   & 1.938(60)  & 1.943(59)  & 1.943(59)  \\
$d_1$       & 0.251(26)  & 0.249(26)  & 0.249(26)  \\
$\log{d_2}$ & -16.70(94) & -16.64(93) & -16.65(93) \\
$d_3$       & 2.83(31)   & 2.84(31)   & 2.84(31)   \\
$\log{C_1}$ & $\star$    & ---        & -13.9(1.1) \\
$\g_1$      & $\star$    & ---        & 2.15(11)   \\
$\log{C_3}$ & -14.2(1.1) & -14.6(1.1) & ---        \\
$\g_3$      & 2.26(15)   & 2.15(10)   & ---        \\
$\log{C_4}$ & -14.1(1.2) & -13.89(95) & -13.53(95) \\
$\g_4$      & 1.94(19)  & 1.968(61)  & 2.003(51)  \\
$\log{C_6}$ & -48(23)    & -64(11)    & -64(11)    \\
$\g_6$      & -4.8(4.7)  & -8.4(2.2)  & -8.4(2.2)  \\
\hline
\end{tabular}
\end{center}
\floatcaption{tab:taste5}{{\it Results of fits to Eqs.~(\ref{Afit}),~(\ref{tsplitA})
and~(\ref{tsplitT}), using all 5 ensembles.  A dash indicates
a parameter that was omitted from the fit.  The fit reported in
the second column includes all parameters, but we do not quote a value for the
parameters $\log{C_1}$ and $\g_1$ (entries indicated by a $\star$ symbol)
since this fit could not resolve them.  See text for further explanation.}
}
\end{table}

\subsection{\label{fitF} Fit including taste splittings}
We now proceed to include the taste splittings, using,
in addition to Eq.~(\ref{Afit}), also Eqs.~(\ref{tsplitA}) and~(\ref{tsplitT}).
Only the masses of pions with tastes corresponding
to the matrices $\G_{i5}$ and $\G_{ij}$ were reported in Ref.~\cite{LSD2},
in addition to the mass of the $\G_5$ pion (the Nambu--Goldstone pion),
limiting us to consider only $\D_A$ and $\D_T$.
With also $M_\p$, $F_\p$ and $M_\t$, this gives us 5 data points per ensemble.

The second column of Table~\ref{tab:taste5} reports the results of the fit
that includes all 13 parameters occurring in tree-level staggered dChPT.
We find $\c^2/{\rm dof}=16.0/12$,
for a $p$-value of 0.19.  The first thing to notice is that the results
for the ``non-taste'' parameters are in very good agreement with
the results of the limited fit shown in Table~\ref{tab:basic}.

While the fit includes all the dChPT parameters, we do not report any value
for the parameters $\log{C_1}$ and $\g_1$.  It turns out that,
in effect, the $\c^2$ function has a flat direction in the subspace
spanned by these two parameters, leaving them undetermined.
In order to understand this situation, consider first
the parameters $\log{C_6}$ and $\g_6$.
The negative mean values found for these parameters imply that
$C_6 E(\g_6) = \exp[\log{C_6}+(4-\g_6)v(m)]$ is negligibly small at the lighter
masses, and becomes significant only for the highest one or two masses.
A similar, but more dramatic, effect occurs in the case of the term
$C_1 E(\g_1)$, which turns out to be significant for the highest mass only.
This means that only one linear combination of $\log{C_1}$ and $\g_1$
(the one defined by the value of $v(m)$ at the largest mass)
was constrained by the data, leaving the
orthogonal linear combination undetermined.

Having seen that the existing data cannot resolve all the taste-splitting
parameters, we also tried fits in which each operator occurring
in Eq.~(\ref{tsplit}) is omitted in turn. The 3rd and 4th columns
of Table~\ref{tab:taste5} report the results of the fits where we have set
to zero $C_1$ or $C_3$, respectively.  Both fits have an equally good $\c^2$
and a $p$-value of 0.14.  The fit with $C_6=0$ has $\c^2=36.8$
while the fit with $C_4=0$ has $\c^2=292$.  Since both of them
have a very low $p$-value, we do not report their results.

It is interesting that, by far, the worst fit is the one where we
have set $C_4=0$.  In other words, the data requires the presence
of the $C_4 E(\g_4)$ term in the fit.  This is nicely consistent
with the taste splittings found in QCD, in the following sense.
Due to the absence of the light scalar,
the pattern of tree-level taste splittings in ordinary ChPT is much simpler,
and corresponds to setting $E(\g_i)=1$ everywhere in Eq.~(\ref{tsplit}) \cite{AB}.
The actual taste splittings exhibit an almost equally spaced spectrum:
the differences $\D_A-\D_P$, $\D_T-\D_A$, $\D_V-\D_T$ and $\D_S-\D_V$
are all roughly constant (independent of the fermion mass)
and equal to each other.  This approximate equality
is explained by the dominance of the $C_4$ term.  As can be seen
in Eq.~(\ref{tsplit}), its coefficient takes on the values 0, 1, 2, 3, and 4.
Hence, in QCD, the constant displacement of the mass squared
between adjacent tastes is given by $C_4$ itself.\footnote{%
  $C_4$ still depends on the bare coupling.  See, \eg, Ref.~\cite{MILC}.
}
Returning to dChPT, the fact that $C_4$
cannot be omitted from the fit shows that, once again, the $C_4$ term
is the most important one.

\begin{table}[t]
\begin{center}
\begin{tabular}{|c|c|c|}
\hline
$\c^2$/dof  & 6.1/9      & 6.2/9      \\
$p$-value   & 0.73       & 0.72       \\
\hline
$\g_*$      & 0.936(19)  & 0.936(19)  \\
$\log{d_0}$   & 1.929(61)  & 1.928(61)  \\
$d_1$       & 0.233(24)  & 0.233(24)  \\
$\log{d_2}$ & -16.18(86) & -16.18(86) \\
$d_3$       & 3.02(32)   & 3.01(32)   \\
$\log{C_1}$ & ---        & -12.3(2.9) \\
$\g_1$      & ---        & 2.49(91)   \\
$\log{C_3}$ & -13.0(3.0) & ---        \\
$\g_3$      & 2.50(92)   & ---        \\
$\log{C_4}$ & -12.7(1.8) & -12.3(1.9) \\
$\g_4$      & 2.16(53)   & 2.22(57)   \\
$\log{C_6}$ & -24(17)    & -25(18)    \\
$\g_6$      & -0.2(3.6)  & -0.3(3.7)  \\
\hline
\end{tabular}
\end{center}
\floatcaption{tab:taste4}{{\it Results of fits to Eqs.~(\ref{Afit}),~(\ref{tsplitA})
and~(\ref{tsplitT}), omitting the $am=0.00889$ ensemble.
As in Table~\ref{tab:taste5}, a dash indicates a parameter
that was omitted from the fit.}
}
\end{table}

\subsection{\label{4ens} Fits with 4 ensembles}
The results reported in the previous subsection mean that some of
the tree-level taste splitting parameters of dChPT acted as nuisance parameters
in our fits:
their presence is required in order to obtain a reasonably good fit,
and yet they remain largely undermined.  We have also explained how,
in effect, these parameters serve to fit the taste-splitting data
at the highest one or two masses, while having a very small,
and often negligible, effect on the quality of the fit for the lighter masses.

This situation motivates us to also consider fits in which the highest mass,
$am=0.00889$, is omitted.  Due to severe numerical instabilities,
we did not attempt a fit with all 13 parameters.  The results of
the fits with $C_1=0$ and with $C_3=0$ are reported in Table~\ref{tab:taste4}.
Both of these fits now have a $p$-value slightly larger than 0.7.
Figure~\ref{fit4F} shows the results obtained for the taste splittings.
(The results for $M_\p$, $F_\p$ and $M_\t$
are visually the same as in Fig.~\ref{fitC}.)
The fit with $C_6=0$ has $\c^2=10.2$, and a $p$-value of $0.49$,
which by itself would be acceptable.  However, it gives rise
to $\log{C_3}=-5(8)$ and $\g_3=4.9(2.5)$, \ie,
these parameters are much less well determined by this fit
than by the fits reported in Table~\ref{tab:taste4}.  Moreover,
we regard the central value of $\g_3$ obtained in this fit as unphysical.
As before, the fit with $C_4=0$ is inconsistent, having $\c^2=117$.

Once again one can see that the values of the 5 ``non-taste'' parameters
are essentially the same as in all previous fits.
As for the taste-splitting parameters, while the mean values are consistent
within error with the 5-ensemble fits, the errors themselves
are significantly larger.  In view of the high $p$-value of the 4-ensemble fits,
the errors reported in Table~\ref{tab:taste4} provide a more realistic
estimate of current uncertainties in the data.

\begin{figure}[t!]
\vspace*{4ex}
\begin{center}
\includegraphics*[width=7cm]{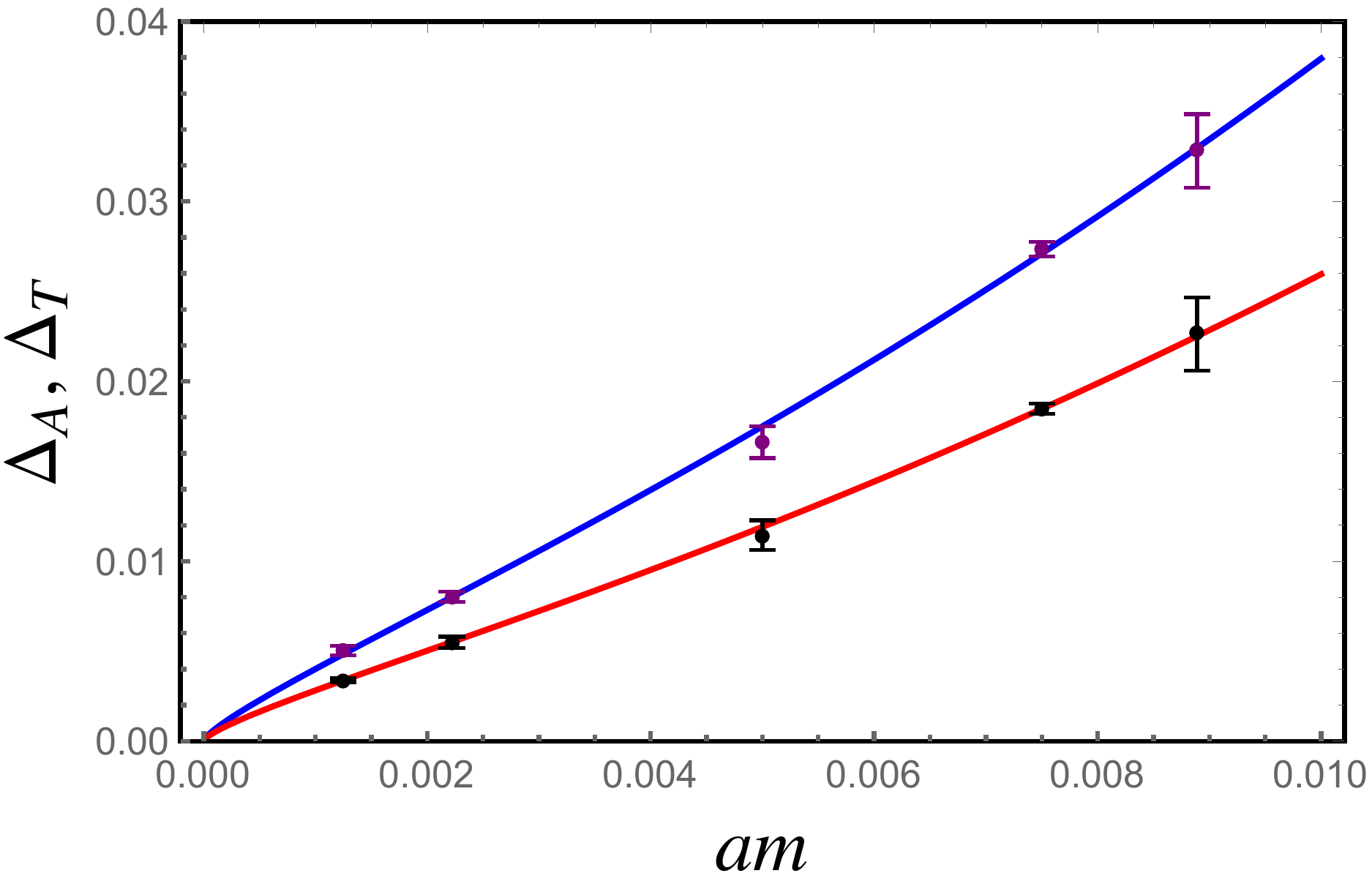}
\hspace{0.5cm}
\includegraphics*[width=7cm]{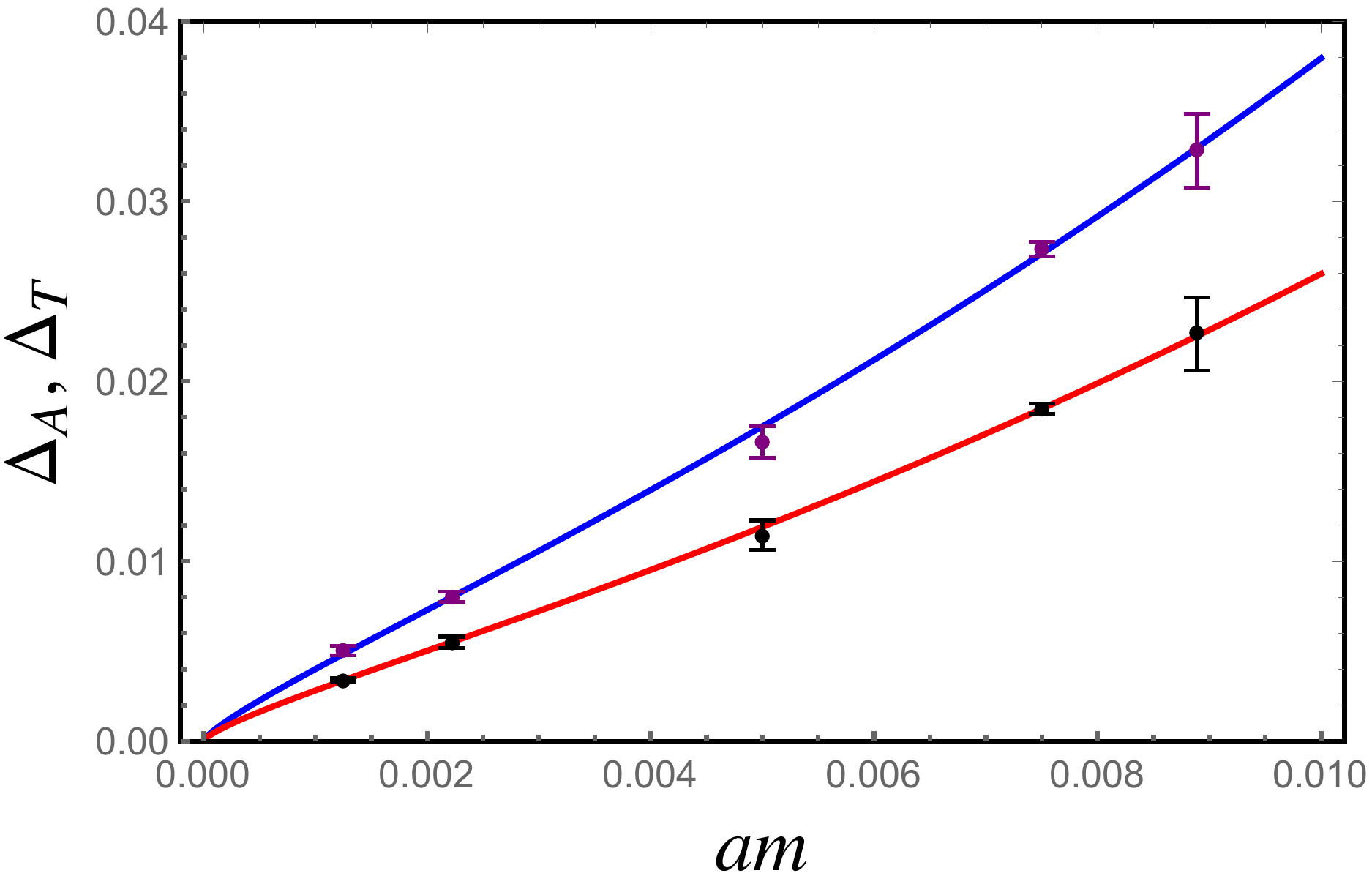}
\end{center}
\begin{quotation}
\floatcaption{fit4F}%
{{\it Representation of the taste splittings found in the 4-ensemble fits
reported in Table~\ref{tab:taste4}. $\D_A$ is plotted in red,
and $\D_T$ in blue. Left: fit with $C_1=0$.  Right: fit with $C_3=0$.}}
\end{quotation}
\vspace*{-4ex}
\end{figure}

\section{\label{conclusion} Conclusion}
In this paper, we studied to what extent tree-level dilaton ChPT
(dChPT)
describes the pseudo-Nambu--Goldstone sector and the light singlet scalar state presented in the lattice data of Ref.~\cite{LSD2}
for the $\SU(3)$ gauge theory with $N_f=8$ fermions in
the fundamental representation.

The simulations reported in Ref.~\cite{LSD2} used staggered
fermions, at 5 different fermion masses and one value of the
gauge coupling, which, in turn, means a single lattice spacing.
We showed that dChPT can be extended to
incorporate the taste-breaking effects that are generally present with staggered fermions, arguing that, therefore,
this provides a nice additional test of the applicability of
dChPT to the data. We fitted data for the
pion mass, the pion decay constant, the dilaton mass,
and the two taste-split pion masses for which Ref.~\cite{LSD2}
provides data.

Even at tree level, staggered dChPT contains quite a few
parameters, 13 in total.    With only 25 data points, it is a
challenge to determine all parameters.
Indeed, attempting to fit all parameters simultaneously we found that
two of them remain undetermined.  The taste-breaking sector contains
four operators, and discarding the pair of parameters associated with
each of these operators in turn we found that some of
the resulting fits are reasonably good when all 5 mass values are included.
As we have explained, the highest-mass ensemble is particularly problematic
when attempting to fit the limited available taste-splitting data.
Omitting this ensemble, we find that the remaining 4 ensembles
are well described by some of the fits in which one of the taste-breaking
operators is omitted.  Should data become available for the two pion tastes
not considered in Ref.~\cite{LSD2}, this would allow fitting also $\D_V$ and $\D_S$
in addition to $\D_A$ and $\D_T$.  This would provide a more stringent test
of dChPT, and may lead to a much better determination of
the taste-splitting parameters.

With these caveats, we believe that dChPT provides a
good explanation for the fermion-mass dependence of the
taste splittings, which in this theory is very different from the
analogous taste splittings in QCD, and for which standard
staggered ChPT has not provided a convincing explanation.
We do not claim that dChPT is the only possible explanation
of these lattice data; it will be very interesting to test
other low-energy approaches proposed in the literature
\cite{BLL,GGS,CM,MY,HLS,AIP1,AIP2,CT,LSDsigma,AIP3},
particularly if they can provide a valid description
of the taste-breaking effects.

For the mass anomalous dimension we found $\g_*=0.936(19)$.
We remind the reader that this value was obtained from data at a
single lattice spacing, and does not include a continuum extrapolation.
This value is in good agreement with the result of Ref.~\cite{Kutietal},
as well as with an earlier, more qualitative, analysis,
based on the eigenmode number \cite{Annaetal}.
An interesting feature is that, in all cases where the anomalous dimensions
$\g_i$ associated with the four-fermion taste-breaking operators were
relatively well determined, their mean values turned out to be
in the range of 1.9 to 2.5, or, in other words, about twice the value
of the mass anomalous dimension.  This result is intuitively appealing,
if we remember that every four-fermion operator contains twice as many
fermion fields as the mass operator.

Our fits also confirm that the simulations of Ref.~\cite{LSD2} are
in the ``large-mass'' regime, in which
the theory shows approximate hyperscaling \cite{largemass}.
In Fig.~\ref{fitC} this is evident
from the near-flatness of the ratios $M_\p^2/F_\p^2$ and
$M_\t^2/F_\p^2$, with the strong downward curvature
predicted by dChPT occurring mostly at smaller values of
$am$ where no data points are available.   Using
\begin{equation}
\label{lhssaddle}
\frac{(3-\g_*)m}{4c_1\cm}=\frac{md_1}{d_2}\sim 4\times 10^6 am\ ,
\end{equation}
which even for the smallest value of $am=0.00125$ is
of order $5\times 10^3$, we conclude that the left-hand
side of Eq.~(\ref{saddle}) is indeed much larger than one for all
masses in Eq.~(\ref{fermionmasses}), thus confirming that these
masses are in the large-mass regime.

A consequence of this is that also the values of $aF_\p$ at the
fermion masses~(\ref{fermionmasses}), which range from
approximately $0.02$ to $0.05$, are much larger than
the chiral-limit value $aF_\p(m=0)=a\hf_\p=0.00049(22)$.
At the smallest fermion mass, the linear spatial volume in
Ref.~\cite{LSD2} is $L/a=64$, leading to $\hf_\p L\approx 0.03$.
This implies that a much larger volume  would be needed to
study the theory in the $\e$-regime.
For comparison, at the smallest fermion mass in the simulation, $am=0.00125$,
one has $F_\p L\approx 1.3$, and $M_\p L\approx 5.3$, so that the
simulations of Ref.~\cite{LSD2} are solidly in the $p$-regime,
and finite-volume corrections are expected to be very small.

Recently, Ref.~\cite{Kutietal20} reported tests of the SU(3) theory with two
sextet (symmetric-representation) Dirac fermions
in the $\e$-regime.  Random matrix theory (RMT) was used
to determine the condensate in the massless limit, finding a value
which is in agreement with
another low-energy description \cite{AIP1} in which the
tree-level dilaton potential takes on a different form from the one
that follows from the power counting underlying dChPT.
We stress, however, that dChPT does not provide
any predictions whatsoever for these simulations,
even if we disregard the fact that the theory under study
contains fermions in a higher representation of the gauge
group.  {The reason is that extrapolation of the $p$-regime data for $F_\p$
to zero fermion mass using dChPT indicates that for the $\e$-regime studies
considered in Ref.~\cite{Kutietal20} one has $F_\p L \ll 1$ for the
chosen combination of volume and fermion mass.
As a result, the small-mass results are
outside the range of applicability of dChPT.}
Were it possible to much enlarge the volume, while keeping $am\approx 0$,
until eventually the condition $F_\p(am\approx 0) L \approx \hf_\p L\approx 1$
would be satisfied,
then, and only then, all measured quantities would have to agree with
the predictions of dChPT,
if indeed dChPT is the correct effective theory at low energy.

We comment that, as discussed in Ref.~\cite{BGKSS}, one can consider
a partially-quenched setup where the sea fermions are kept in the $p$-regime
and only the valence fermions are in the $\e$-regime.
In such a setup the dilaton expectation value is determine
only by the mass of the sea fermions.  This setup can provide for limited $\e$-regime tests of dChPT
on currently available ensembles.

\vspace{2ex}
\noindent {\bf Acknowledgments}
\vspace{2ex}

We thank the LSD collaboration for providing the full covariance
matrix for the spectroscopic data reported in Ref.~\cite{LSD2}, and we
thank Julius Kuti for discussions.
MG's work is supported by the U.S. Department of
Energy, Office of Science, Office of High Energy Physics, under Award
Number DE-SC0013682.  EN's work is supported by the U.S. Department of
Energy, Office of Science, Office of High Energy Physics, under Award
Number DE-SC0010005.
YS is supported by the Israel Science Foundation
under grant no.~491/17.

\appendix
\begin{boldmath}
\section{\label{$t_0$} The $m$ dependence of $\sqrt{8t_0}$}
\end{boldmath}
In QCD, $t_0$ is implicitly determined by the equation
\cite{MLflow}
\begin{equation}
\label{sett0}
t_0^2 \svev{E(t_0,x)} = 0.3 \equiv c_0 \ .
\end{equation}
(In \SU($N$) gauge theories with $N\ne 3$, one needs an appropriate
rescaling of the constant $c_0$, see for example Ref.~\cite{TDflow}.)
Here
\begin{equation}
\label{defE}
E(t,x) = \frac{1}{4}\, G_{\m\n}^a(t,x) G_{\m\n}^a(t,x) \ ,
\end{equation}
where $G_{\m\n}^a(t,x)$ is the field strength of the flow field, which is
subject to the boundary condition $G_{\m\n}^a(0,x)=F_{\m\n}^a(x)$,
where $F_{\m\n}^a(x)$ is the field strength of the dynamical field,
and with the convention that the classical action is
$\frac{1}{4g^2}\int d^4x\, F_{\m\n}^aF_{\m\n}^a$.

In QCD, $E(t)$ admits a chiral expansion,
from which it follows that \cite{BG}
\begin{equation}
\label{Em}
\svev{E(t,x)} = const. + O(m) + O(m^2\log(m)) \ .
\end{equation}
The $m$-independence of the leading term,
and the (related) fact that logarithmic corrections occur only at NNLO,
feeds into the chiral expansion for $t_0$,
making it a particularly convenient quantity
for scale setting in QCD.

Let us now consider a nearly conformal, confining theory.
On dimensional grounds, the leading (operator) expression for $E(t,x)$ in
the effective theory is
\begin{equation}
\label{Eteff}
E(t,x) = C_0(t) e^{4\t}\ ,
\end{equation}
with $\t$ is the dilaton field, and
$C_0(t)$ an unknown function of $t$.   Hence
\begin{equation}
\label{Eteffvev}
\svev{E(t,x)} = C_0(t) e^{4v}\ ,
\end{equation}
where $v=v(m)$ is the classical solution determined by Eq.~(\ref{saddle}).  We see that, unlike in QCD,
now the leading term depends on the fermion mass, because $v(m)$ does.
Differentiating the relation
\begin{equation}
\label{sett0LM}
t_0^2 C_0(t_0) e^{4v(m)} = c_0 \ ,
\end{equation}
and using the approximate solution~(\ref{vapprox}) valid for large $\frac{m}{c_1 \cm}$, gives
\begin{equation}
\label{dt0dm}
\frac{d\log(t_0)}{d\log(m)} \ \sim \ -\frac{2}{1+\g_*}
\frac{1}{1+\half\frac{\partial\log(C_0)}{\partial\log(t_0)}}\ .
\end{equation}
Naive hyperscaling, as determined on dimensional grounds,
would suggest that the right-hand side of Eq.~(\ref{dt0dm}) be equal to $-\frac{2}{1+\g_*}$.
The presence of the correcting factor on the right-hand side
implies that $t_0$ does not have to obey the same hyperscaling relation
as hadron masses and decay constants do in the large-mass regime
\cite{largemass}.

Moreover, the leading-order dependence of $\svev{E(t)}$ on $m$,
coupled with our ignorance about the functional form of $C_0(t)$,
implies that, unlike in QCD, we cannot write down a useful chiral expansion
for $t_0$ in a nearly conformal theory.  In a way, the situation is similar
to that of the Sommer scale; one can empirically parametrize the
(unknown) dependence of the Sommer scale on the quark mass.  But, as for the Sommer scale, we do not have a theory-driven explicit expression to back up a particular
expansion for the mass dependence.

\begin{figure}[t]
\vspace*{4ex}
\begin{center}
\includegraphics*[width=10cm]{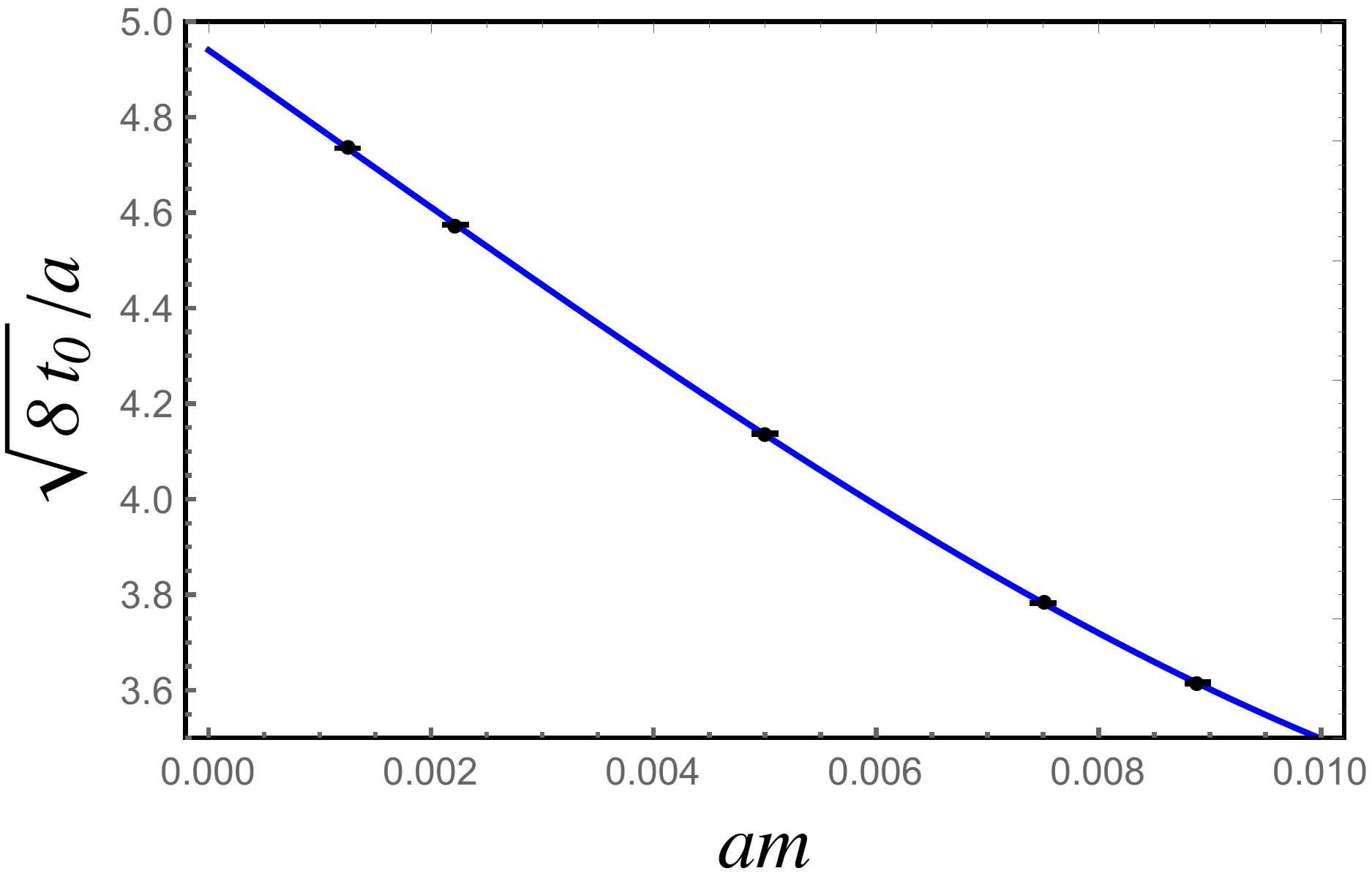}
\end{center}
\begin{quotation}
\floatcaption{t0fit}%
{{\it Fit of data for $\sqrt{8t_0}/a$ from Ref.~\cite{LSD2} to a cubic
polynomial.}}
\end{quotation}
\vspace*{-4ex}
\end{figure}

It still interesting to fit the $am$ dependence of $\sqrt{8t_0}/a$,
for which Ref.~\cite{LSD2} also reported results with very small errors.
Such a fit is purely ``phenomenological,'' because of
the lack of a dChPT prediction for this dependence.   We
find that a fit of the data of Ref.~\cite{LSD2} to a cubic polynomial in $am$,
\begin{equation}
\label{cubic}
R(am)=\sum_{n=0}^3 a_n (am)^n
\end{equation}
yields a statistically successful fit.
This fit yields $\c^2_{min}=0.78$, with one degree of freedom;
the parameter values are given by
\begin{eqnarray}
\label{Rpar}
a_0&=&4.9400(66)\ ,\\
a_1&=&-164.0(6.1)\ ,\nonumber\\
a_2&=&-8(14)\times 10^2\ ,\nonumber\\
a_3&=&277(92)\times 10^3\ .\nonumber
\end{eqnarray}
Figure~\ref{t0fit} shows the fit.  The value for $a_0$ is of
interest, because it provides an estimate of $\sqrt{8t_0}/a$
in the chiral limit.

\vspace{3ex}

\end{document}